\documentclass[prc,aps,preprint,showpacs]{revtex4}
\usepackage{graphicx}

\begin{document}
\title{Oblate deformation of light neutron-rich even-even nuclei} 

\author{ Ikuko Hamamoto$^{1,2}$ }

\affiliation{
$^{1}$ {\it Riken Nishina Center, Wako, Saitama 351-0198, Japan } \\ 
$^{2}$ {\it Division of Mathematical Physics, Lund Institute of Technology 
at the University of Lund, Lund, Sweden} }   




\begin{abstract}
Light neutron-rich even-even nuclei, of which the ground state is 
oblately deformed, 
are looked for, examining the Nilsson diagram based on realistic 
Woods-Saxon potentials. 
One-particle energies of the Nilsson diagram are calculated 
by solving the coupled
differential equations obtained from the Schr\"{o}dinger equation in coordinate
space with the proper asymptotic behavior for $r \rightarrow \infty$ 
for both 
one-particle bound and resonant levels.
The eigenphase formalism is used in the calculation of one-particle resonant
energies.
Large energy gaps on the oblate side of the Nilsson diagrams are found 
to be related
to the magic numbers for the oblate deformation of the harmonic-oscillator
potential where the frequency 
ratios ($\omega_{\perp} : \omega_{z}$) are simple rational numbers. 
In contrast, for the prolate deformation the magic numbers obtained from simple 
rational ratios of ($\omega_{\perp} : \omega_{z}$) of the harmonic-oscillator
potential are hardly related to the particle numbers, at which 
large energy gaps 
appear in the Nilsson diagrams based on realistic 
Woods-Saxon potentials.  
The argument for an oblate shape of $^{42}_{14}$Si$_{28}$ is given.  
Among light nuclei the nucleus $^{20}_{6}$C$_{14}$ is found 
to be a good candidate for having the
oblate ground state.  In the region of the mass number $A \approx 70$ 
the oblate ground state may be found in the nuclei around  
$^{76}_{28}$Ni$_{48}$ in addition to $^{64}_{28}$Ni$_{36}$.  

\end{abstract}


\maketitle

\newpage

\section{INTRODUCTION} 
The ground states of some nuclei are described by densities and mean fields 
that 
are spherical, while others are deformed.    
Quadrupole deformation is by far the most important type of deformation, 
and the features of most deformed nuclei are consistent with 
an axially symmetric shape.
Prolate shape is absolutely dominant in axially-symmetric 
quadrupole-deformed nuclei 
observed so far, 
while in the absence of 
pair correlation one obtains the number of prolate systems equal to that of
oblate ones in the simple models such as harmonic-oscillator 
and single-j-shell.    
A simple intuitive reason for the absolute dominance of prolate shape 
in observed deformed nuclei 
is not yet clarified \cite{HM09}.
Thus, at present one may say that the shell-structure which strongly favors   
oblate shape is needed for the realization of oblate nuclei.  

Recently the neutron-rich nucleus $^{42}_{14}$Si$_{28}$ is reported to show 
a rotational spectrum \cite{ST13}, namely 
the ratio of observed excitation energies is  
$E(4^{+})/E(2^{+}) = 2.93(5)$ where
$E(I^{\pi})$ expresses the excitation energy of the lowest level with
the spin-parity $I^{\pi}$, though the
spin-assignment of the $4^{+}$ state is not yet pinned down experimentally.  
The observed value of E($2^{+}$) is not small enough, namely  
the relatively low moment of inertia indicates an oblate deformation  
though a decisive experimental evidence for the oblate
shape is not yet obtained.   
One may wonder the reason why the nucleus with the neutron-number $N$=28 
is deformed 
and not spherical, 
as the particle number 28 is a well-known magic-number 
in the j-j coupling shell-model 
 \cite{MGM49, HJS49}.  
 
When I look for even-even nuclei in the range of $6 \leq Z \leq 40$, 
of which the observed electric quadrupole moment 
of the first-excited 2$^+$ state is clearly positive corresponding to 
an oblate shape (or a fluctuation towords oblate shape), 
I find \cite{RAG89,NJS05} only the following five 
nuclei: 
$^{12}_{6}$C$_{6}$, $^{28}_{14}$Si$_{14}$, $^{34}_{16}$S$_{18}$, 
$^{36}_{18}$Ar$_{18}$ and $^{64}_{28}$Ni$_{36}$.  
In addition, the ground state of the proton-rich $N$=$Z$ nucleus 
$^{72}_{36}$Kr$_{36}$ seems to 
have most likely an oblate shape \cite{AG05}.  

Examining the values of proton- and neutron-numbers of the oblate nuclei 
mentioned above reminds me of the magic numbers of oblate deformations with
the frequency ratio $(\omega_{\perp} : \omega_{3})$ = (1:2) and (2:3) 
in the deformed harmonic-oscillator (h.o.) potential 
(see Figure 6-48 of \cite{BM75}). Namely, those magic numbers are 
$Z$=$N$= 6, 14 , 26, 44, 68,... 
for the (1:2) deformation and $N$=$Z$= 6, 8, 14, 18, 28, 34, 48, 58,... 
for the (2:3) deformation.  
When the neutron- and/or proton-numbers are equal to one of 
those magic numbers, the system 
in the deformed h.o. potential  
is supposed to be especially stable for respective 
($\omega_{\perp} : \omega_{z}$) deformations, though the
total deformation is determined by both proton- and neutron-numbers.    
In the deformed h.o. model 
those magic numbers and the degeneracy of 
one-particle levels in the shells above and below can 
be understood in terms of a simple intuitive picture \cite{BM75}.
In reality, besides the difference between the h.o. potential and realistic
potentials, one should take into account that 
the neutron 
shell-structure around the Fermi level may drastically change
in nuclei towards neutron drip line \cite{AO00,IH07,IH12}.   

In the present paper the possible presence of the oblate shape of the
ground state of lighter neutron-rich even-even nuclei is explored, 
studying the shell-structure of 
the Nilsson diagram (namely
neutron one-particle spectra as a function of axially-symmetric 
quadrupole deformation) 
based on realistic deformed Woods-Saxon potentials.  
The deviation from (the similarity to) the magic numbers of 
the deformed h.o. potential is discussed for 
prolate (oblate) shape.  

In Sec. II the brief summary of the model used is given, while numerical
results, the simple interpretation and the direct consequences are presented 
in Sec. III.  Conclusions and discussions are given in Sec. IV.

\section{MODEL}
The basic points of the model used in the present paper are very similar to  
those of Ref. \cite{IH12}.  
While weakly-bound neutrons in neutron-rich nuclei close to the neutron drip
line make a contribution especially to the tail of the self-consistent nuclear
potentials, the major part of the nuclear potential for neutrons 
is provided by protons, which are deeply bound in the case of
neutron-rich nuclei.  
And, since the effective interaction to be used in the
Hartree-Fock calculations of neutron-rich nuclei
far away from the stability line is not yet well pinned down, in the present
study I use Woods-Saxon
potentials with the standard parameters which are described 
on p.239 of Ref. \cite{BM69}.  

The calculation of energies of bound particles is an eigenvalue problem 
solving the coupled differential equations
derived from the Schr\"{o}dinger equation in coordinate space, together
with the correct asymptotic behavior of bound 
wave-functions in respective channels
for $r \rightarrow \infty$.  
In contast, one-particle resonant energies are obtained  
by solving the coupled differential equations in coordinate space using
the asymptotic behavior of scattering states in respective channels and are  
defined as the energies, at which one of the
eigenphases increases through $\pi /2$ as the energy increases 
\cite{RGN66,IH05}. 

One-particle resonance is absent if none of the eigenphases increase through
$\pi /2$ as the energy increases.  As the energy increases, the width of a given
resonant level becomes generally larger, 
and finally at a certain energy the one-particle
resonant level with a given $\Omega^{\pi}$ (or $\ell j$ in spherical cases) 
is no longer obtained, where $\Omega$ expresses the angular-momentum component
of the particle along the symmetry axis.
The disappearance of those 
one-particle resonant levels can occur just above the Fermi level of 
some neutron drip line
nuclei.  See, for example, the Nilsson diagram for neutrons of 
$^{20}_{6}$C$_{14}$ shown in Fig. 2.  
For simplicity, the
calculated widths of one-particle resonant levels are not always given, 
because the widths are not of major interest in the present work.  
When the energy gap in one-particle spectra of nuclei
close to the neutron drip line is dicussed, 
it is important to explore the shell structure 
including one-particle resonant levels.   

I use the knowledge that a large energy gap in one-particle spectra 
for some deformation in the Nilsson
diagram leads to the stability of the system for the deformation and the 
particle-number. 
The deformation parameter 
$\beta$ of the ground
state obtained from experimental data is relatively large for lighter
nuclei, but certainly smaller than $|\beta| < 0.6$.  Therefore, in the following
I study the deformation
region of $|\beta| < 0.6$.

\section{NUMERICAL RESULTS}
It is known that for light stable nuclei the realistic Nilsson diagrams 
for protons 
and neutrons are not so different.    The Nilsson diagram for 
well-bound protons has been studied in various publications, and even 
the diagrams based on modified-oscillator potentials can be safely 
used to
find out the dependence of the shell structure on the proton number and
deformations.  
Whether a given nucleus is deformed or not
depends on both proton- and neutron-numbers, and the oblate
ground-state of even-even
nuclei is rather rare. Thus, when possible lighter neutron-rich
oblate nuclei
are searched for, the proton numbers are fixed to be 
those that are known to be favored by oblate shape.  

The volume conservation is taken into account in Nilsson diagrams 
shown in the present paper, while it is neglected in those of 
previous publications \cite{IH07, IH12}.  The neglect hardly affects 
the discussion of the shell-structure change in neutron-rich nuclei.

\subsection{Oblate shape of the nucleus $^{42}_{14}$Si$_{28}$}
It is known that a large energy gap appears at the proton number $Z$=14  
on the oblate side of the Nilsson diagram obtained from 
both the modified-oscillator 
potential and any realistic Woods-Saxon potential appropriate for stable 
nuclei.  The particle number 14 is
also the magic number for both the (1:2) and (2:3) oblate deformations in the
h.o. potential.

In Fig. 1 the Nilsson diagram for neutrons of $^{42}_{14}$Si$_{28}$ 
is shown.  
The energy distance $\varepsilon(2p_{3/2}) - \varepsilon(1f_{7/2})$ 
is only 1.65 MeV, which is much smaller than the $N$=28 energy gap
known in the j-j coupling shell-model.
That means, the spherical shape 
is not particularly favored by $^{42}$Si.
It is in fact known (for example, 
see Fig. 2 in Ref. \cite{IH12}) that the $2p_{3/2}$ and $1f_{7/2}$ levels come 
very close to 
each other (or the level order can be even reversed), when those two levels
become very weakly bound or low-energy resonant. 
In contrast, from both Fig. 1 of the present paper 
and Fig. 2 of Ref. \cite{IH12} it is seen that
a large energy gap is developed at $N$=28 on the oblate side.  The gap 
corresponds exactly to 
the magic number 28 for the (2:3) oblate deformation of the 
h.o. potential shown in Fig. 6-48 of Ref. \cite{BM75}.     
Namely, at $N$ = 28  
on the oblate side of Fig. 1 
four doubly-degenerate Nilsson levels with the h.o. principal
quantum-number $N_{ho}$ = 3 are occupied,
while none of  $N_{ho}$ = 2 Nilsson orbits are occupied.  (Note that 
one-particle levels of the h.o. model 
in Fig. 6-48 of Ref. \cite{BM75} are classified 
by the quantum numbers, ($n_{\perp}, n_{z}$), therefore, 
the one-particle levels  
are not always doubly-degenerate but have the degeneracy of 
$2 (n_{\perp} + 1)$.)   
In short, the possible oblate deformation of the nucleus $^{42}$Si can be
understood as a result of the combination of the facts: 
(i) the disappearance of 
spherical $N$ = 28 magic number due to the shell-structure change 
in very neutron-rich
nuclei; (ii) the $N$ = 28 remains as ''a magic number'' for
the moderate-size ($\beta \approx -0.4$) oblate deformation in the same way 
as the magic number for the (2:3) deformation of the h.o. 
potential; 
(iii) the oblate shape is much favored by $Z$ = 14.  

Other energy gaps of a considerable size found on the oblate side of Fig. 1 
for 
$\beta \approx -0.4$ are $N$ = 14 and 18, which are also the magic numbers 
for the oblate (2:3) deformation of the h.o. potential.  
In contrast, it is noticed that the neutron numbers corresponding to the 
energy gaps of an appreciable size 
 that are seen 
on the prolate side of Fig. 1 (for example,
$N$ = 16, 24, and 28) do not correspond to the magic numbers for 
the prolate
(3:2) deformation of the h.o. potential.    

The different correspondence between the realistic
Woods-Saxon potential and the h.o. potential for the prolate
deformation from the  
oblate deformation  
seems to come mainly from the different behavior of the Nilsson one-particle 
levels connected 
to the high-j shell  
(the $1f_{7/2}$ shell in the present case) on the
prolate side from the oblate side.  
The different behavior of the high-j Nilsson levels 
was discussed in detail in Ref. \cite{HM09} 
in relation to the numbers of oblate/prolate nuclei.  
On the oblate side the shell-structure coming from 
the unique presence of the high-j orbits in
realistic potentials is disturbed soon after deformation sets in. In contrast,  
on the prolate side the shell-structure due to the large spin-obit splitting 
is kept in the range of the realistic
quadrupole deformation and, thus, the particle number, at which 
a large energy gap occurrs, is considerably 
different from that of 
the h.o. potential.

\subsection{Possible oblate shape of the nucleus $^{20}_{6}$C$_{14}$}
One may try to find out good candidates for 
other light neutron-rich nuclei with an oblate shape.  
There is a large energy gap above the proton number $Z$ = 6 on the oblate side
of the Nilsson diagram with realistic potentials as well as the 
deformed modified-oscillator potential. For example, see Fig. 5-1 of Ref.
\cite{BM75}.
Therefore, a promissing candidate is $^{20}_{6}$C$_{14}$. 
From the measured values, $E(2_1^+) \approx 1.6$ MeV \cite{MS08,MP11} 
and $B(E2;2_1^+ \rightarrow 0^+) \approx$ 7.5 $e^2 fm^4$ \cite{MP11}, the nucleus
$^{20}$C is most probably deformed.  
The occupation of  
the nearly degenerate $2s_{1/2}$ and $1d_{5/2}$ levels by several neutrons 
may well induce a
deformation due to the nuclear Jahn-Teller effect \cite{IH07, IH12}.  
One question is whether or not the large energy gap at $N$ = 14 remains  
for the oblate deformation with $\beta \approx -0.4$,     
when those two 
single-particle levels, $2s_{1/2}$ and $1d_{5/2}$, become very weakly-bound. 
In Fig. 2 the Nilsson diagram for neutrons of $^{20}$C is shown.   
The width of the $\Omega^{\pi} = 1/2^{+}$ level connected to the  
$2s_{1/2}$ level becomes extremely large immediately above zero
energy, and the level cannot survive as a resonant state for 
$\beta < -0.215$.  
In essence, there is no $\Omega^{\pi} = 1/2^{+}$ one-particle resonant 
level for 
$-0.6 < \beta < -0.2$, of  
which the wave-function has an appreciable probability inside the nucleus, 
and, thus, 
a large energy gap appears at $N$=14.  
Therefore, the nucleus $^{20}_{6}$C$_{14}$ is  
a promising 
candidate for a light neutron-rich nucleus with an oblate shape. 

In Fig. 2 large energy gaps on the oblate side 
are observed at $N$= 6, 8, and 14.  
Those
neutron numbers are the magic numbers for the (2:3) oblate deformation of the
h.o. potential.  The numbers of the orbits with a given $N_{ho}$ 
which lie below 
respective particle numbers are again exactly the same as those 
numbers for the (2:3) deformation.

\subsection{Oblate neutron-rich nuclei in the $A \approx 70$ region}
In this slightly heavier mass region 
the one-particle level-density increases and 
the pair correlation may become more important, which is neglected in the
present paper.  Even in such cases, a low one-particle level-density at some
deformation for a given particle-number in Nilsson diagrams 
usually plays a role in determining the
deformation of the ground state of the system.  

Examining the shell-structure around $\beta \approx -0.4$ in Fig. 3 of Ref. 
\cite{IH12}, large energy gaps are found at $N$=36 and 48.   
The neutron numbers $N$=34 and 48 are the magic numbers 
for the (2:3) deformation 
of the h.o. potential. 
The $N$=48 gap in Fig. 3 of Ref. \cite{IH12} corresponds exactly to 
the magic number 48 
for the (2:3)
deformation in the sense that five doubly-degenerate 
$N_{ho}$=4 levels are occupied while one doubly-degenerate $N_{ho}$=3 level 
is unoccupied.  
The occurrence of the large energy gap at $N$=36    
instead of the magic number $N$=34 for the (2:3) deformation 
comes from the
occupation of the lowest-lying doubly-degenerate $\Omega^{\pi}$ = $9/2^{+}$ 
level, which
originates from the $1g_{9/2}$ level and is lowered steeply 
as oblate deformation increases.  
The wave function of the $\Omega^{\pi}$ = $9/2^{+}$ level is almost pure
$1g_{9/2}$ in the present range of $\beta$-values. 
The large energy gap at $N$ = 36 instead of $N$ = 34 
comes from the presence of 
the high-j-shell orbit ($1g_{9/2}$ orbit) 
around the Fermi level in this neutron-number region, which
is strongly pushed down by the large spin-orbit splitting that is absent in
the h.o. model.  

From the shell-structure in Fig. 3 of Ref. \cite{IH12}, 
one may note that possible candidates for 
the oblate ground-state of neutron-rich even-even nuclei 
in the $A \approx 70$ region    
are the nuclei around $^{76}_{28}$Ni$_{48}$ in addition to   
$^{64}_{28}$Ni$_{36}$.   

As already pointed out in subsection III. A, it is seen from Fig. 3 of Ref.
\cite{IH12} that 
none of the large energy gaps for the prolate deformation with $\beta \approx
0.4$ correspond to 
the magic numbers for the (3:2) deformation of the h.o.  
potential,
$N$= ,,, 22, 26, 34, 46, 54,,, .

\section{CONCLUSIONS AND DISCUSSIONS}
The possible oblate ground-state of light neutron-rich even-even  
nuclei is searched for, using the shell structure appearing in  
the Nilsson diagrams based on the Woods-Saxon
potential with standard parameters.  For nuclei close to the neutron-drip-line
the shell-structure including one-particle resonant levels is studied.  

It is shown that the neutron-numbers, 
at which large energy gaps appear for some
oblate deformations of Nilsson diagrams based on the realistic potentials, 
can be easily related to the 
magic numbers coming from the degeneracy of the energy 
$\hbar (n_{\perp} \omega_{\perp} + n_z \omega_{z})$ for the oblate deformation 
of the h.o. potential, in which the ratios of
frequencies  
($\omega_{\perp} : \omega_z$) are simple rational numbers.
Because of this simple property of the shell-structure on the oblate side of 
the Nilsson diagram, 
which is common
to the Woods-Saxon potentials with realistic parameters, 
one may pretty reliably predict the light nuclei, of which
the ground state may have an oblate shape.  
A good candidate for the oblate shape of very light nuclei is $^{20}$C.
An oblate shape can be
expected for $^{42}_{14}$Si$_{28}$, partly because $N$=28  is not really 
a magic number 
for spherical shape of the neutron-rich nucleus, instead, it is a magic number 
for the (2:3) oblate deformation of the h.o. 
potential and partly because $Z$=14 is a magic number for the (2:3) (and
also (1:2)) oblate deformation.  Indeed, one observes a large energy gap both 
at $N$=28 and $Z$=14 for the moderate-size (say, $\beta \approx -0.4$) 
deformation 
in the Nilsson diagram based on realistic Woods-Saxon potentials.
In the mass-number $A \approx 70$ region the possible candidates 
for oblate shape  
are the nuclei around 
$^{76}_{28}$Ni$_{48}$ in addition to $^{64}_{28}$Ni$_{36}$.  

In contrast, on the prolate side of the Nilsson diagram based on 
realistic Woods-Saxon potentials large energy gaps 
are hardly found at the particle numbers, 
which correspond to the magic numbers with a 
simple rational frequency-ratio ($\omega_{\perp} : \omega_{z}$) 
of the h.o. potential.   
In other words, the particle numbers, at which large energy gaps appear on the
oblate side of the Nilsson diagram, are quite common to various realistic
potentials, while on the prolate side they may depend on some details of
respective one-body potentials.

An oblate ground-state 
of $^{42}$Si was also obtained in available shell-model calculations \cite{FN09,
YU12} using various effective interactions with and without tensor forces.
In those shell-model calculations the h.o. wave-functions are always
used and, therefore, the shell-structure change in neutron-rich nuclei 
due to the unique behavior of weakly-bound (and/or resonant) 
low-$\ell$ neutrons is absent. 
The shell-structure change may possibly be mimicked 
by adjusting some part of effective interactions together with input
one-particle energies.   
On the other hand, the shape of the system is the property of the one-body mean
field.  It is the author's opinion that the explanation of the oblate shape of
$^{42}$Si given in the present 
paper is simple, intuitive and widely applicable.   

The author is grateful to Professor K. Matsuyanagi for useful discussions 
which she had in the initial stage of the present work.

\newpage

\noindent
{\bf\large Figure captions}\\
\begin{description}
\item[{\rm Figure 1 :}]
Calculated one-particle energies for neutrons of $^{42}_{14}$Si$_{28}$ 
as a function of axially-symmetric quadrupole deformation.
Bound one-particle energies at $\beta$ = 0 are $-$8.50, $-$7.24, $-$2.42,and
$-$0.77 MeV for the $2s_{1/2}$, $1d_{3/2}$, $1f_{7/2}$, and $2p_{3/2}$ levels, 
respectively, while
one-particle resonant $2p_{1/2}$ and $1f_{5/2}$ levels are obtained at +0.20 MeV
with the width of 0.20 MeV and
+3.25 MeV with the width of 0.58 MeV, respectively, which are denoted by  
filled circles.  The resonant $\Omega^{\pi}$ = $1/2^{-}$ level 
which is connected to the 
resonant $2p_{1/2}$ level at $\beta$ = 0 cannot be obtained for 
$\beta < -0.19$ 
corresponding to $\varepsilon_{\Omega} > 1.42$ MeV.  
One-particle
resonant energies for $\beta \neq$ 0 are not plotted unless they are important 
for the present discussion.  
For simplicity, calculated widths of one-particle resonant levels are not shown.
The neutron numbers, 14, 18, 20 and 28, which
are obtained by filling all lower-lying levels, are indicated with circles.  
One-particle levels with $\Omega$ = 1/2, 3/2, 5/2 and 7/2  are expressed 
by solid, dotted, long-dashed and dot-dashed curves, respectively, 
for both positive and negative parities.  The parity of levels can be seen from
the $\ell$-values denoted at $\beta$ = 0; $\pi$ = $(-1)^{\ell}$.

\end{description}

\begin{description}
\item[{\rm Figure 2 :}]
Calculated one-particle energies for neutrons of $^{20}_{6}$C$_{14}$ 
as a function of axially-symmetric quadrupole deformation.
Bound one-particle energies at $\beta$ = 0 are $-$7.66,
$-$1.28 and $-$0.71 MeV for the $1p_{1/2}$, $1d_{5/2}$ and 
$2s_{1/2}$ levels, respectively, while one-particle resonant $1d_{3/2}$ level 
is obtained at +3.57 MeV with the width of 3.02 MeV.
The resonant $\Omega^{\pi}$ = $1/2^{+}$ level which is connected to the 
bound $2s_{1/2}$ level at $\beta$ = 0 cannot be obtained for $\beta < -0.215$ 
corresponding to $\varepsilon_{\Omega} > 0.38$ MeV.  
On the other hand, the resonant $\Omega^{\pi}$ = 1/2$^{+}$ level connected to 
the resonant $1d_{3/2}$ level at
$\beta$ = 0 cannot be obtained for $\beta > 0.35$ corresponding to 
$\varepsilon < 1.54$ MeV, because in the deformation region 
the $s_{1/2}$ component  
becomes appreciable in the wave function.  For smaller $\beta$ values 
the
major component of the wave function of the $\Omega^{\pi}$ = 1/2$^{+}$ level 
is $1d_{3/2}$ and, thus, the resonant level survives.   
The neutron numbers, 6, 8 and 14, which are obtained by filling all lower-lying
levels, are indicated with circles. 
One-particle levels with $\Omega$ = 1/2, 3/2 and 5/2 are expressed by solid,
dotted and long-dashed curves, respectively, for both positive and negative
parities.  
 
\end{description}

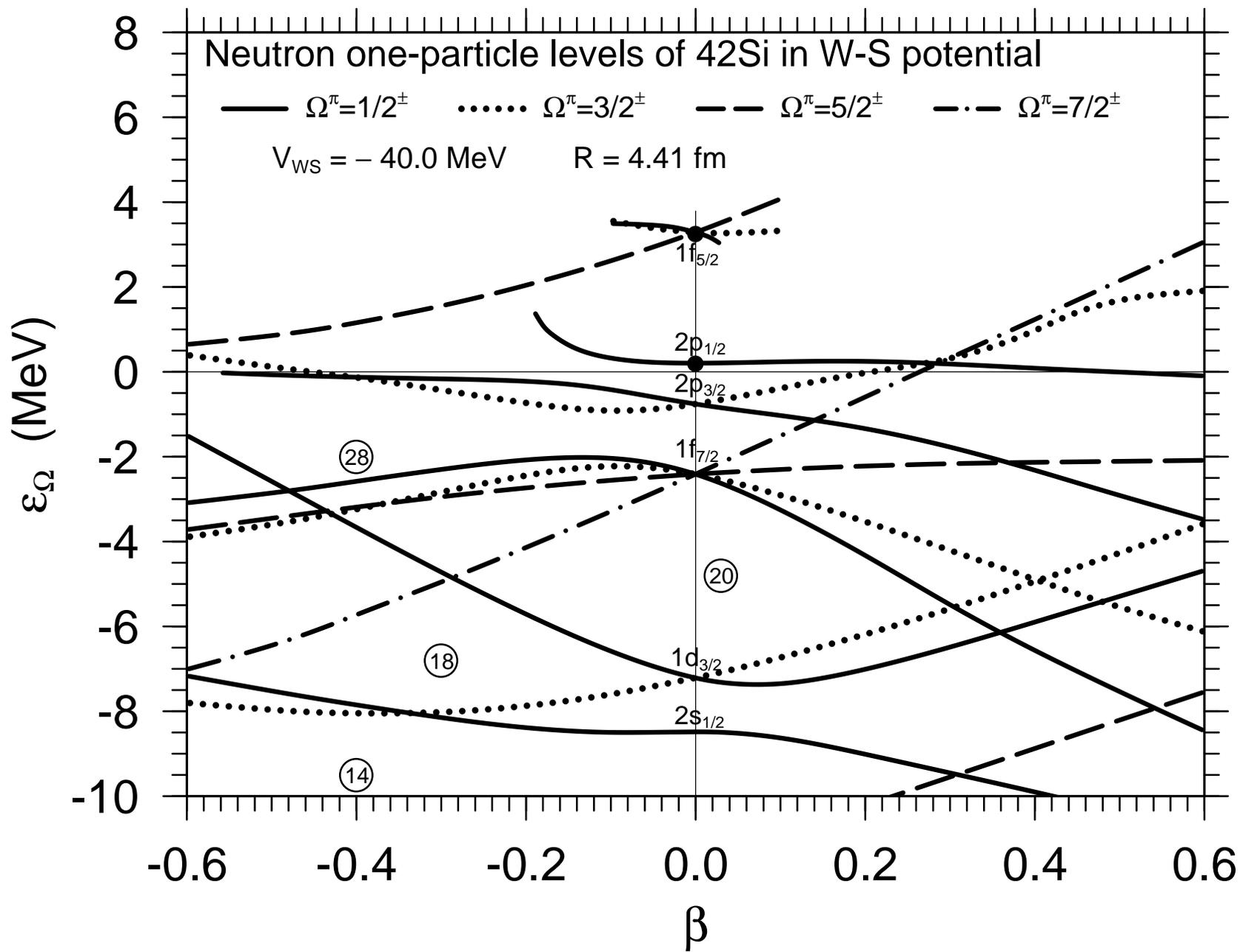
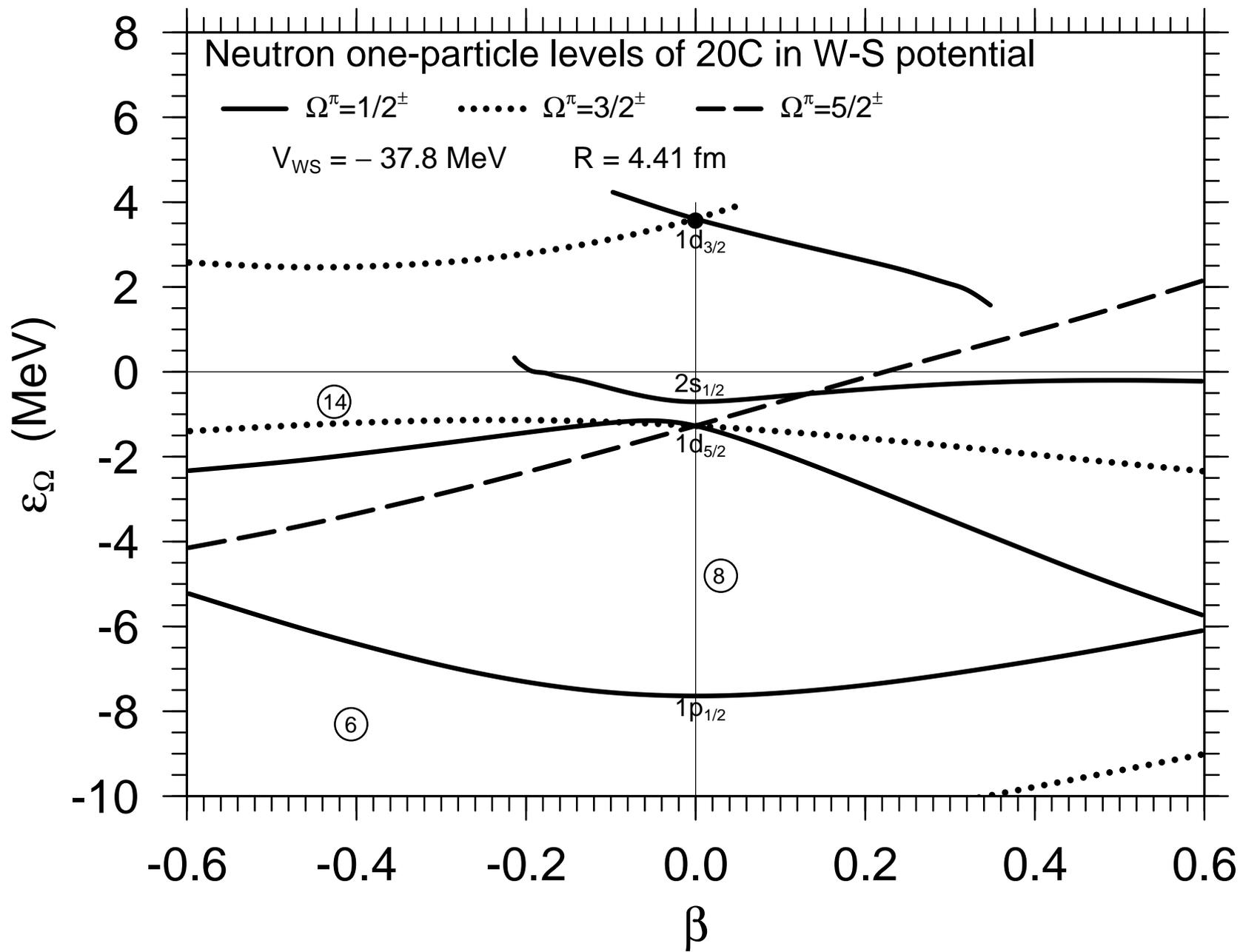
\end{document}